\documentclass[letterpaper,twocolumn,10pt]{article}
\usepackage{usenix2019_v3}

\usepackage{tikz}
\usepackage{amsmath}
\usepackage{caption}
\usepackage{subcaption}
\usepackage{xcolor}
\usepackage{tabularx}
\usepackage{multirow}
\usepackage{amsmath}
\usepackage{mathtools}
\usepackage{float}
\usepackage{filecontents}

\begin{document}

\date{}

\title{\Large \bf RAPID-Serve: Resource-efficient and Accelerated P/D Intra-GPU Disaggregation}

\author{
{\rm Amna Masood}\\
\texttt{amna.masood@amd.com}\\
Advanced Micro Devices
\and
{\rm Pratishtha Gaur}\thanks{Work done while interning at AMD Research and Advanced Development}\\
\texttt{pratishtha.gaur@amd.com}\\
Advanced Micro Devices
\and
{\rm Nuwan Jayasena}\\
\texttt{nuwan.jayasena@amd.com}\\
Advanced Micro Devices
} 

\maketitle

\newcommand{\nj}[1]{\textcolor{green}{NJ: #1}}
\newcommand{\am}[1]{\textcolor{purple}{AM: #1}}

\begin{abstract}

Two widely adopted techniques for LLM inference serving systems today are hybrid batching and disaggregated serving. A hybrid batch combines prefill and decode tokens of different requests in the same batch to improve resource utilization and throughput at the cost of increased latency per token. In contrast, disaggregated serving decouples compute-bound prefill and bandwidth-bound decode phases to optimize for service level objectives (SLOs) at the cost of resource under-utilization and KV-cache transfer overheads. To address the limitations of these techniques, we propose RAPID-Serve: a technique to concurrently execute prefill and decode on the same GPU(s) to meet latency SLOs while maintaining high throughput and efficient resource utilization. Furthermore, we propose Adaptive Resource Management for runtime compute resource allocation, optionally leveraging CU masking (a fine-grained Compute Unit partitioning feature on AMD Instinct\textsuperscript{TM} GPUs). RAPID-Serve provides up to 4.1x (average 1.7x) unconstrained throughput improvement and 32x and higher (average 4.9x) throughput improvement under SLO constraints, showing it as an effective strategy compared to the state-of-the-art approaches, particularly in resource-constrained environments.

\end{abstract}

\section{Introduction}
Large language models (LLMs) are increasingly deployed in interactive applications like coding assistants, chatbots, and realtime decision tools. Prior work on interactive LLM serving\cite{nanoflow2024} 
emphasizes that inter-token latency (ITL) of LLM applications should remain within the range of human reading speed (approximately 150-300ms per token) for user experience. However,
emerging techniques like test-time scaling, chain of thought reasoning and multi-step planning require generation of significant number of internal 'thinking' tokens which users never see. In such scenarios, the latency criterion shifts from how fast the users can read tokens to how efficiently the system can process large amounts of intermediate generation.
As a result, optimizing for human-perceived streaming speed alone is no longer sufficient; system design must increasingly focus on high-throughput token generation, while simultaneously improving GPU resource utilization, since internal reasoning can easily account for the majority of both compute demand and memory pressure. 
This shift places demands on LLM serving systems to deliver lower inter-token latency, high throughput, and efficient use of limited accelerator resources.

In LLM inference, request execution consists of two main phases: prefill, which processes the input prompt and initializes the KV-cache, and decode, which generates output tokens auto-regressively. 
These two stages exhibit different performance bottlenecks.
In practice, serving systems rely on two widely used approaches to balance latency and throughput: hybrid batching and disaggregated serving. Hybrid batching\cite{yu2022orca} 
improves overall utilization by combining prefill and decode tokens into a single execution batch. However, this coupling inflates inter-token latency for ongoing requests, making it difficult to satisfy strict ITL SLOs. Chunked prefill techniques\cite{agrawal2024taming}\cite{splitwise2024} mitigate delays by slicing long prefills into smaller segments, yet decode latency is still amplified due to prefill and decode progressing in lockstep.
Disaggregated serving attempts to remove inter-phase interference entirely by executing prefill and decode on separate hardware. However, achieving an optimal balance between throughput and latency depends on several design factors, including model size, architecture, traffic patterns, and bandwidth sensitivity. Consequently, practical adoption of disaggregated serving is still limited by the complexity of the optimization space and the need for system-level coordination \cite{mitra2025beyondbuzz}.


For latency-sensitive workloads, these constraints make it difficult for existing systems to achieve both low latency and high throughput at the same time, especially in resource-constrained environments.

Our key insight is that predictable inter-token latency and high throughput can be achieved simultaneously by allowing prefill and decode to make progress concurrently on each GPU. 
This creates new opportunities to overlap the two phases without introducing coordination overheads or KV-cache transfers. At a high level, enabling controlled intra-GPU prefill-decode (P/D) concurrency breaks the strict phase coupling of hybrid batching that inflates latency, while still retaining the resource efficiency of shared execution.

To realize this insight, RAPID-Serve runs prefill and decode phases of different requests concurrently without combining them into a single batch, so each phase can make independent progress. Depending on the current workload, these phases may be assigned distinct compute resources to minimize interference and adhere to latency SLOs, while maintaining high throughput. To adapt to varying workload intensity, the system dynamically adjusts the amount of compute allocated to each phase.
This flexible allocation model avoids the rigid constraints of traditional disaggregation designs, eliminates KV-cache transfer overheads, and maintains batching efficiency, 
significantly improving overall serving performance and resource utilization.

This paper makes the following contributions:
\begin{enumerate}
\item Introduces a new serving engine called RAPID-Serve that overlaps prefill and decode on the same device(s).
    \item Analyzes the performance characteristics of existing serving solutions and quantifies overheads due to P/D coupling and KV cache transfer.
    \item Provides an analysis of distinct performance characteristics and requirements of different phases of LLM inference.
    
    \item Introduces a profiling-driven resource allocation technique 
    to select configurations that meet latency SLOs with minimal resource usage.
    
    \item Evaluates the proposed solution and shows that RAPID-Serve delivers low latency and high throughput across diverse workloads for practical LLM serving deployments.
\end{enumerate}

\section{Background}
\subsection{LLM Inference}
LLM inference has two phases with contrasting resource demands. The prefill phase processes the entire input prompt to produce the first output token. Because all prompt tokens can be evaluated in parallel, prefill is dominated by large compute-intensive matrix multiplications and self-attention operations. This phase typically exhibits high arithmetic intensity and benefits from wide parallelism across the GPU’s compute units. The resulting key–value (KV) entries generated during prefill are stored in KV cache for use in subsequent token generation.

The decode phase is typically autoregressive, generating one output token at a time.
Unlike prefill, decode operates sequentially, performing self-attention over the accumulated KV cache for each new token. This shifts the workload towards matrix–vector operations with lower arithmetic intensity and higher sensitivity to memory bandwidth and cache reuse. Even when decode requests are batched, operations such as attention still require large KV cache reads per request. 

The KV cache is a major contributor to decode-time memory consumption and system pressure. Its size grows linearly with both batch size and sequence length, and also depends on the model architecture (e.g., number of layers, hidden size, number of attention heads, and head sharing strategy). For models using Multi-Head Attention (MHA), Multi-Query Attention (MQA) or Grouped-Query Attention (GQA), the per-request KV cache size can be approximated as:
\begin{equation}
    \text{KV cache bytes = }2 \cdot L \cdot S \cdot H \cdot D \cdot E
\end{equation}
where L is the number of layers, S is the sequence length, H is the number of KV heads, D is the per-head hidden dimension, and E is the number of bytes per element determined by the numeric precision. The factor of two accounts for storing both keys and values.

The distinct performance characteristics of the two phases create different optimization goals, making it challenging to design a serving system that efficiently optimizes both phases simultaneously.


\begin{figure}[H]
    \begin{center}
    \includegraphics[width=1\linewidth,height=7cm]{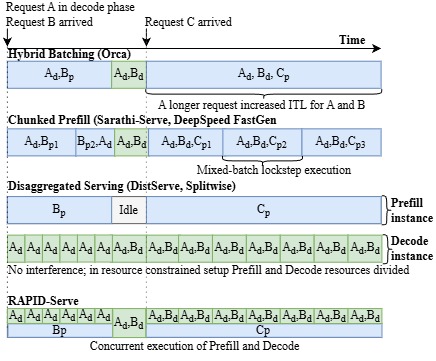}
    \caption{Scheduling approaches. Consider a scenario where request A is in decode phase when B arrives; $B_p$ denotes prefill of request B and $A_d$ denotes decode of request A and , and $B_{p1}$ represents the first chunk of B’s prefill when chunked prefill is applied.
    }
    \label{fig:scheduling}
    \end{center}
\end{figure}
\subsection{Hybrid Batching}
One of the early LLM serving solutions, Orca \cite{yu2022orca}, introduced iteration-level batching, where prefill and decode tokens from multiple requests are combined into a single batch for one model iteration.
Each iteration generates tokens for each active request. 
Prior to Orca, serving systems such as vLLM typically batched only decode phases, while the prefill for each request was executed independently upon arrival—effectively prefill-prioritizing. FasterTransformer\cite{fastertransformer} similarly prioritized decode batches.

Subsequent work, including Sarathi~\cite{agrawal2024taming} and DeepSpeed-FastGen~\cite{deepspeedfastgen2024}, identified and addressed the issues with prefill/decode stalling
problem inherent in the aforementioned scheduling strategies. In decode-prioritizing systems, prefill operations are delayed; in prefill-prioritizing systems, decode steps stall under high request arrival rates. In Orca, because prefill and decode are batched together, a long-running prefill request can increase the ITL of all decode requests in the batch. As shown in Figure \ref{fig:scheduling}, the long prefill of C delays decode requests ${A_d}$ and ${B_d.}$ Sarathi addressed this issue using a piggybacking strategy. They define a maximum per-iteration token budget: each batch is first filled with all active decode tokens, and the remaining capacity is allocated to prefill tokens. If a prefill request exceeds the available budget, its prompt is divided into multiple pieces called "chunks." To support chunking, Sarathi introduces a KV cache management mechanism that stores intermediate chunk results as KV cache entries (Figure \ref{fig:scheduling}). Although this approach mitigates stalls, it still suffers from interference between prefill and decode phases. In decode-heavy workloads, prefill requests may be fragmented into many fine-grained chunks, incurring additional KV cache load/store operations. Conversely, in prefill-heavy workloads with small decode batches, decode tokens are frequently co-batched with prefills, which increases decode ITL. This ITL degradation can range from 2× to 30× \cite{chen2025_disaggregated_inference_18mo_later}.
Medha \cite{agrawal2024medha} introduces adaptive prefill chunk sizing to improve GPU utilization and enable fine-grained preemption for long-context requests. However, because prefills and decodes still share the same execution cycles, adaptive chunking cannot provide stable or low decode ITL
due to high interference in case of long prompts or when the adaptive policy inflates chunk size under load.
Additional analysis is presented in Section~\ref{chunking_overhead}.
RAPID-Serve
breaks the lockstep dependence between prefill and decode, minimizing ITL while still maintaining high throughput.
\subsection{Disaggregated Serving}
\begin{figure*}[htbp]
    \centering
    \begin{subfigure}[t]{0.3\textwidth}
    \centering
    \includegraphics[height=5cm,width=!]{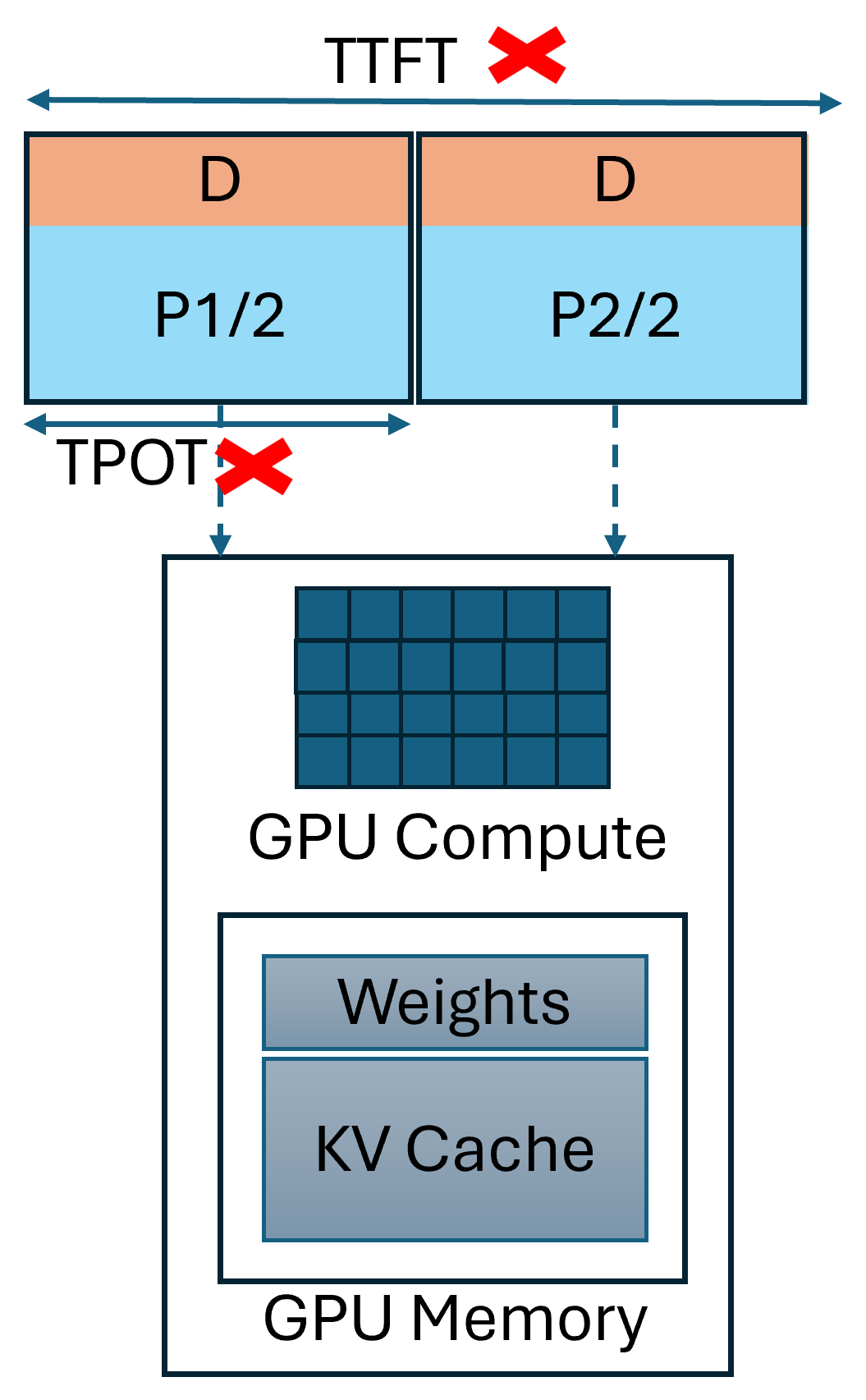}
    \caption{Hybrid Batching with Chunked Prefill}
    \label{fig2a}
    \end{subfigure}
    \hfill
    \begin{subfigure}[t]{0.3\textwidth}
    \centering
    \includegraphics[height=5cm,width=!]{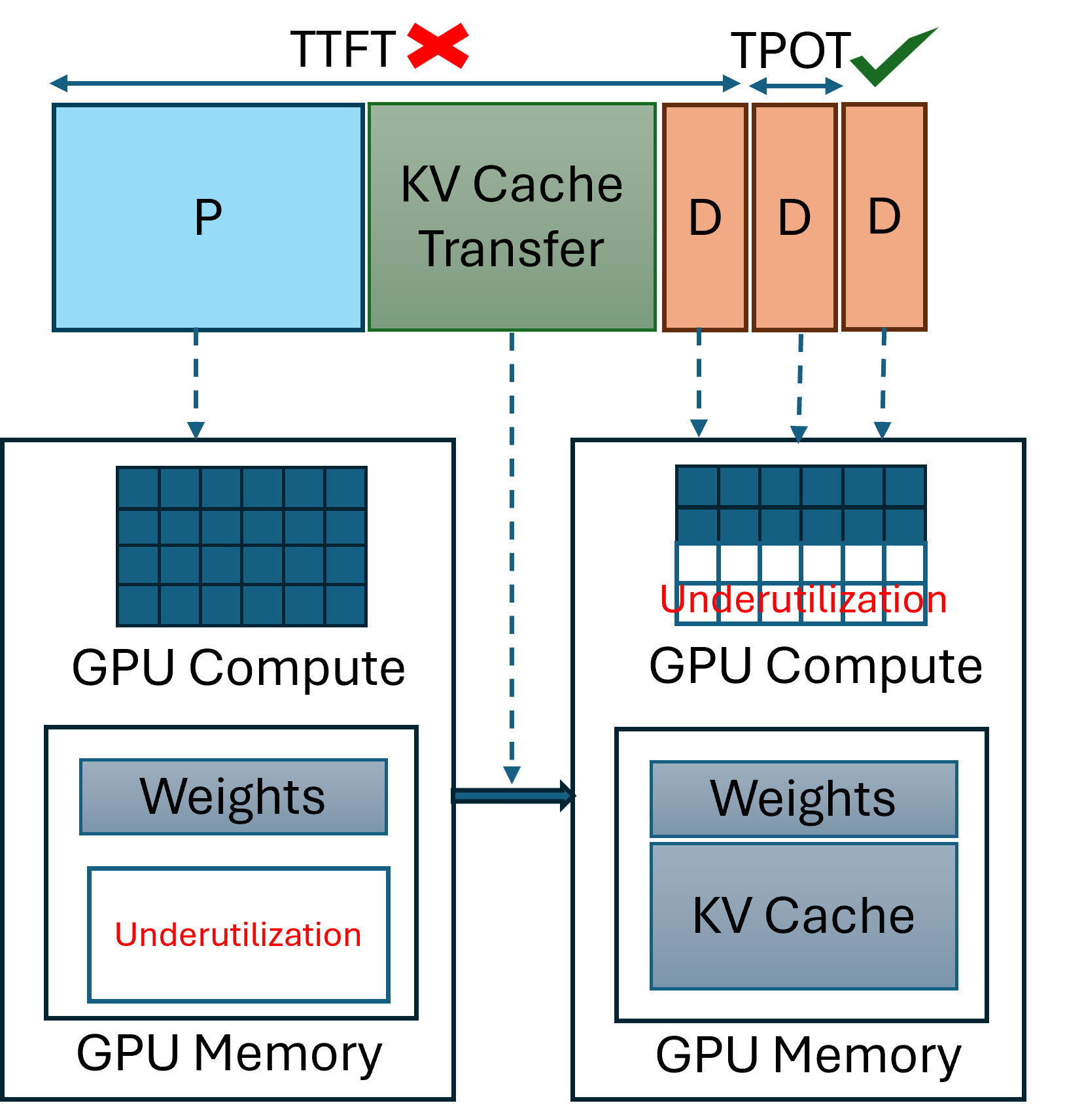}
    \caption{Disaggregated Serving}
    \label{fig2b}
    \hfill
    \end{subfigure}
    \begin{subfigure}[t]{0.3\textwidth}
    \centering
    \includegraphics[height=5cm,width=!]{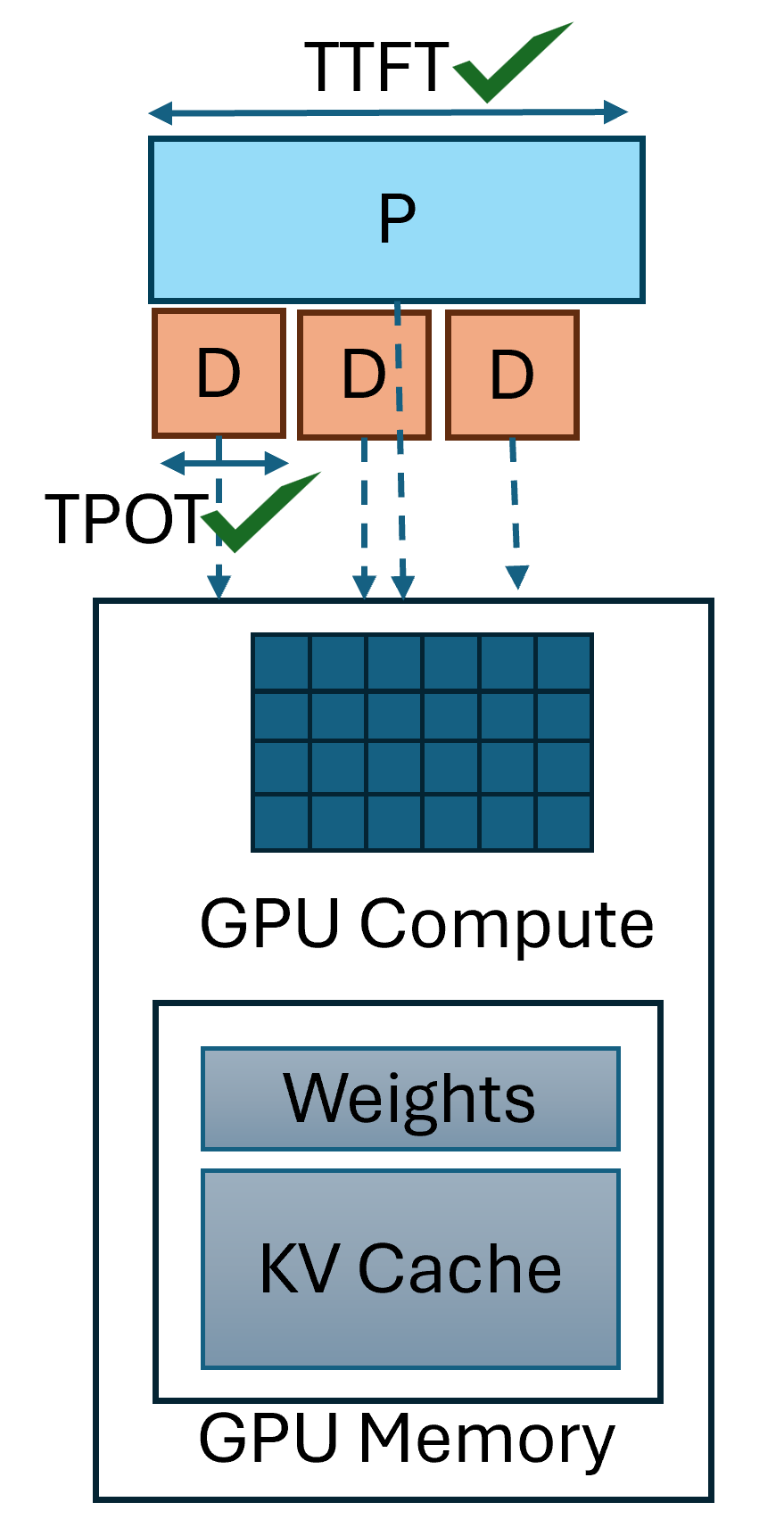}
    \caption{RAPID-Serve}
    \label{fig2c}
    \end{subfigure}
    \caption{Different Serving Techniques
    }
    \label{fig2}
\end{figure*}

To mitigate interference between the prefill and decode phases, recent works~\cite{splitwise2024,zhong2024distserve} propose Disaggregated Serving, where prefill and decode run on separate pools of workers instead of being co-located on the same device(s) as depicted in Figure \ref{fig2b}. Incoming requests are first routed to a prefill instance that performs the compute-intensive forward pass over the entire prompt and materializes the initial KV cache entries. The resulting KV cache is then transferred over the network/interconnect to a decode instance, which then performs token generation over several iterations. This design removes direct resource contention between the two phases and allows each cluster to be provisioned and scaled according to its requirements. However, disaggregation turns the KV cache into a producer–consumer artifact that must be moved across machines on the critical path. Subsequent work focuses on reducing the cost of KV cache movement. Systems such as Mooncake\cite{Qin2025Mooncake} and LMCache\cite{Cheng2025LMCache} explore techniques like hierarchical cache placement, compressed representations, and selective eviction policies to minimize data transfer overheads.

These designs, however, implicitly assume a moderately sized cluster with ample GPUs and high-bandwidth interconnects, where dedicating separate prefill and decode pools is feasible and the additional network hops can be amortized over many requests. In resource-constrained deployments with only one to a few nodes
, disaggregating prefill and decode can be counterproductive \cite{mitra2025beyondbuzz}, 
as the benefits of disaggregation diminish for smaller models and under generation-heavy workloads. Moreover, its effectiveness is fundamentally constrained by interconnect and hardware bandwidth, as the cost of transferring the KV cache between the prefill and decode stages must not become the bottleneck in the overall inference latency. 
\section{Design Considerations}
\subsection{Overheads of Hybrid Batching} \label{chunking_overhead}

Combining prefill and decode requests in a single batch leads to tradeoffs between throughput and latencies (Figure~\ref{fig2a}). To maintain high throughput, larger chunk sizes are preferred at the cost of higher per-iteration latency, which directly corresponds to higher ITL. On the other hand, stringent ITL SLOs necessitate smaller chunk sizes, reducing effective GPU utilization and throughput. Our experiments show that a chunk size of 1K tokens achieves 20\% better throughput on average, albeit with a 30\% increase in ITL, compared to chunk size 512. This tradeoff requires careful orchestration of chunk sizes based on workload.

\subsection{Overheads of Disaggregated Serving} \label{disagg_overhead}
Disaggregated serving has been suggested as an alternative to hybrid batching, allowing for prefill and decode phases to execute on separate compute instances that are optimized for the respective phases. However, given a fixed set of resources, disaggregated serving exhibits a number of overheads compared to aggregated serving approaches.

\subsubsection{KV Transfer} \label{kvtransfer_overhead}
In a disaggregated serving system, an incoming request is first routed to the prefill instance for prompt processing and initial KV cache population. The request and associated KV-cache data is then routed to decode instance for generating the output tokens. The overhead of KV cache transfer depends on several factors, such as the interconnect or network bandwidth, size of KV cache, the prefill/decode instance configurations, etc. Evaluating vLLM's disaggregated serving on a node of 8 AMD Instinct\textsuperscript{TM} MI300X GPUs (4P-4D), KV transfer is shown to have average overheads of 1.4x for throughput and 1.9x for Time-To-First-Token (TTFT). These overheads are typically larger for smaller prompts as there is less opportunity to overlap the transfer with the prompt computation.
Note that in vLLM v1, the first token is calculated again in the decode instance after receiving the KV cache; hence, the KV transfer overhead impacts the generation of the first token rather than the second.
\begin{figure*}[htbp]
    \centering
    \begin{subfigure}[t]{0.48\linewidth}
    \includegraphics[width=0.9\linewidth]{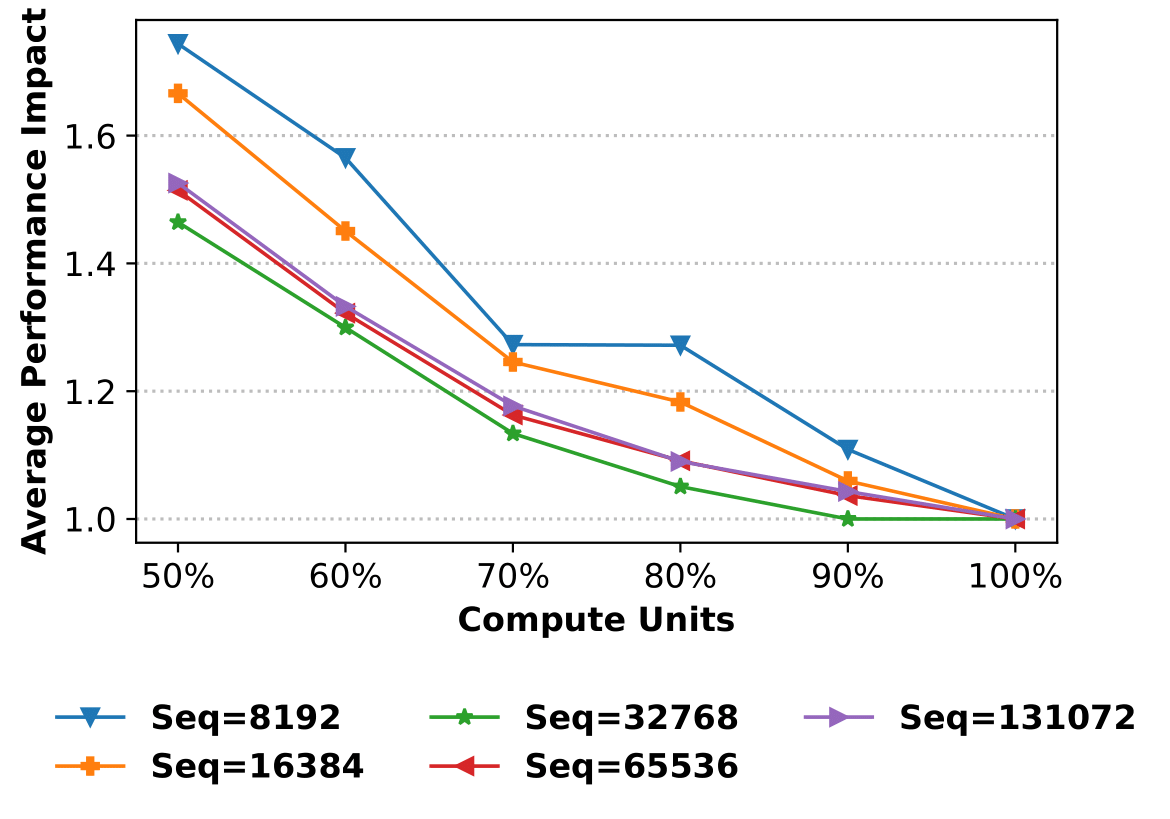}
    \caption{Compute-bound prefill phase exhibits direct correlation of performance with amount of compute resources}
    \label{fig1a}
    \end{subfigure}
    \hfill
    \begin{subfigure}[t]{0.48\linewidth}
    \includegraphics[width=0.9\linewidth]{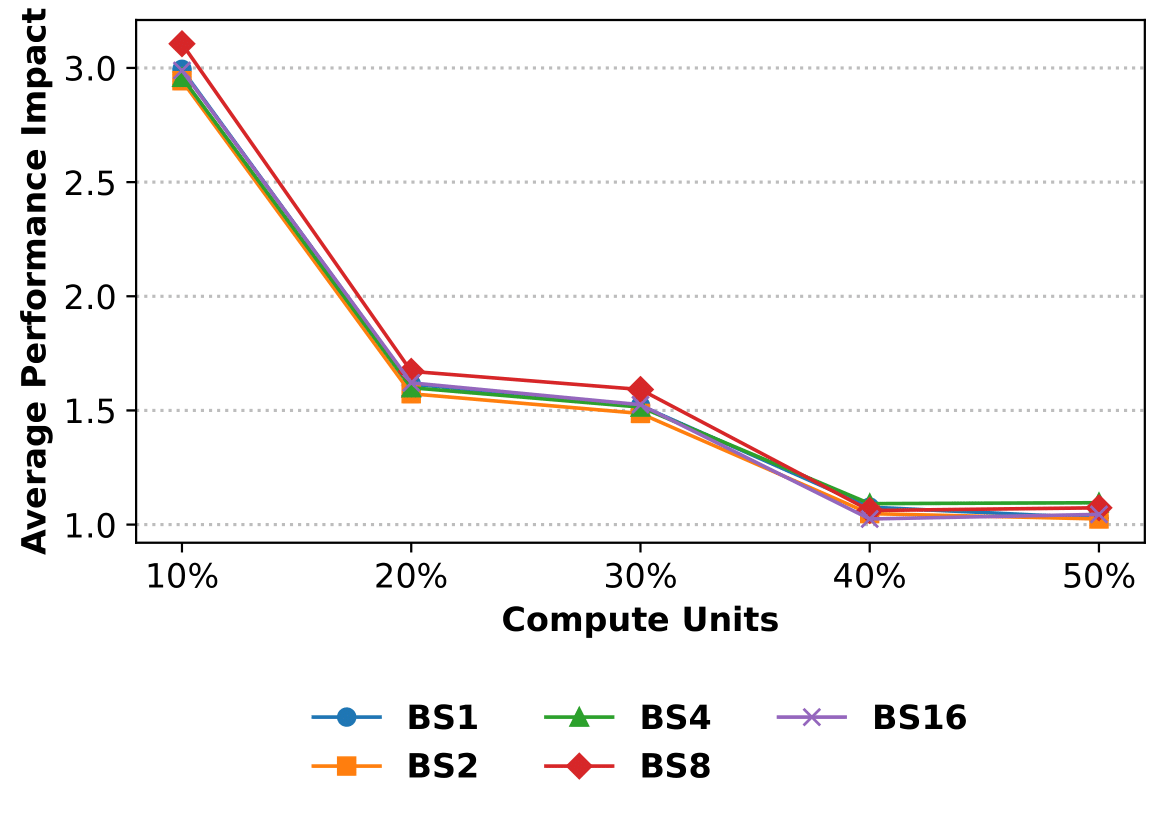}
    \caption{Memory-bandwidth bound decode achieves similar performance with 40-50\% of compute (BS=Batchsize)}
    \label{fig1b}
    \end{subfigure}
    \caption{Resource requirements for inference phases. Y-axis represents performance degradation normalized to performance at 100\% compute units (lower is better, 1=peak performance).
    }
    \label{fig1}
\end{figure*}

\subsubsection{Memory Underutilization} \label{memutil_imbalance}
After the transfer, the KV cache data resides in the memory of decode instance during the longer decode phase, freeing up prefill instance memory. This creates an imbalance in the memory utilization of the instances, where decode instance(s) contain more data for longer time,
whereas prefill instances' memory capacity is underutilized. In a simple case where both prefill and decode are assigned the same number of similar GPUs, this imbalance implies up to 50\% of memory capacity underutilization and consequently, in case of memory-capacity limited inference workload, up to 50\% throughput cost, since half the memory capacity means half the number of requests executing in a single instance.

\subsection{Phase Specific Resource Requirements} \label{resource}
The two phases of LLM inference exhibit distinct performance characteristics: the prefill phase that processes the input prompt is compute-bound, while the generation or decode phase is generally memory bandwidth bound. However, GPUs today have massive compute capabilities along with high memory capacity and bandwidth. Compute-bound workloads do not fully utilize memory-bandwidth, and decode phase does not need all the compute resources. RAPID-Serve proposes to concurrently execute prefill and decode in order to fully utilize the GPU's capabilities.

Figure~\ref{fig1} shows the impact of apportioning compute resources to prefill and decode. For this experiment, a matrix multiplication kernel corresponding to the up projection layer of LlaMA 3.1 70B model's transformer layer was executed on AMD Instint\textsuperscript{TM} MI300X with smaller input sizes corresponding to decode and larger input sizes corresponding to prefill.
The kernels were then allocated specific compute resources using CU masking: a feature for fine-grained allocation of compute units (Section~\ref{cu_masking}).
Figure~\ref{fig1a} shows that the performance of prefill workload decreases as the amount of compute resources decreases, confirming that prefill kernels are compute-bound. On the other hand, decode kernels (Figure~\ref{fig1b}) can achieve similar performance even when the compute resources are halved, showing that compute resources are underutilized during memory-bandwidth bound decode. Concurrent execution of these two phases presents a viable solution to better utilize GPU resources.  


\subsection{Memory Subsystem Interference}
Another form of interference when running multiple kernels concurrently on the same device is due to contention in the memory subsystem. While CU masking enables distinct compute resource allocation to concurrently running kernels, such allocation is not possible for the memory subsystem that includes caches, on-chip network, and memory channels and controllers. Contention for these resources causes memory interference. However, this interference is expected not to significantly impact the performance of compute-bound kernels. Our experiments based on attention kernels show that prefill kernels only exhibit a performance impact of up to 2\% due to memory interference, which is quite modest and acceptable. The memory interference exhibited by decode kernels ranges from 2-5\% on average. 
Repeating the same experiment with GEMM kernels, we observe larger memory interference for decode kernels. However, there is no available mechanism on AMD Instinct\textsuperscript{TM} GPUs to partition the memory resources between concurrent kernels without virtualizing the GPUs. Hence, RAPID-Serve does not mitigate performance degradation due to memory subsystem. 



\section{RAPID-Serve Implementation}
RAPID-Serve is designed to address the limitations of existing serving techniques.  The implementation is based on vLLM v1 (v0.10.2rc3). This section details the implementation choices.
\subsection{Overview} \label{design_overview}
The intuition behind the design of RAPID-Serve is to exploit the advantages of both hybrid chunking -- concurrent execution of P/D for high resource utilization -- and disaggregated serving -- decoupling of the two phases to meet latency SLOs -- on the same device(s). 
Our design goals are as follows:
\begin{enumerate}
    \item Maintain high GPU utilization for compute and memory
    \item Minimize gaps in GPU execution and maintain high concurrency for prefill and decode 
    \item Meet latency SLOs for both prefill and decode
    \item Maintain the advantages of hybrid chunking and disaggregated serving while mitigating their overheads
\end{enumerate}

As described in Section~\ref{chunking_overhead}, while hybrid batching provides high throughput, its main disadvantage is the coupling of decode and prefill requests by combining both in a single batch, meaning that prefill and decode computation is done in a lockstep, causing latencies to exceed strict SLOs 
(Figure~\ref{fig2a}). RAPID-Serve addresses this limitation by executing prefill and decode phases in parallel but not combining them in a single batch. This means that prefill and decode are executed in separate processes and GPU kernels, but the kernels overlap their execution. RAPID-Serve removes the lockstep execution of hybrid batching while still providing the throughput advantage 
(Figure~\ref{fig2c}).

RAPID-Serve inherently addresses the overhead of KV cache transfer associated with disaggregated serving (Section~\ref{kvtransfer_overhead}) by executing prefill and decode concurrently on the same device (design goal \#1) and sharing KV cache between them, thus avoiding any inter-device KV cache transfers. Moreover, as KV cache resides on the same device as prefill and decode, there is no imbalance in memory capacity utilization unlike disaggregated serving (Figure~\ref{fig2}).

\subsection{CU Masking} \label{cu_masking}
RAPID-Serve's design leverages a hardware mechanism exposed by AMD Instinct\textsuperscript{TM} GPUs called CU masking, which allows kernel execution to be restricted to a subset of compute units. 
When CU masking is enabled, there is a bit mask associated with each command queue with each bit corresponding to a specific CU. When a kernel is enqueued on a CU-masked command queue, its workgroups are only executed on the specific CUs.
This feature is transparent to the application: no kernel code changes are required, and the masked-out CUs are simply invisible to the workload. Importantly, CU masking provides spatial control over GPU compute resources and is orthogonal to other optimization mechanisms such as kernel priority or GPU virtualization.
\subsection{Multiprocessing vs. Multithreading}
While LLM inference is generally a GPU-bound task, enabling GPU concurrency also requires that the host CPU tasks also run in parallel. The CPU is responsible for tasks such as handling incoming requests and outgoing responses, scheduling requests for GPU processing, post-processing the outputs and managing the KV cache. Serialization in CPU work can cause gaps in GPU execution and limit concurrency (Figure~\ref{naive}). 
As the use of the Global Interpreter Lock (GIL) in Python limits concurrent execution of multiple threads, RAPID-Serve takes a multiprocessing approach where prefill and decode are executed by different processes to allow true concurrent execution of CPU tasks and avoid interference from different threads (design goal \#2). 
\begin{figure*}[htbp]
    \begin{center}
    \includegraphics[width=0.9\textwidth]{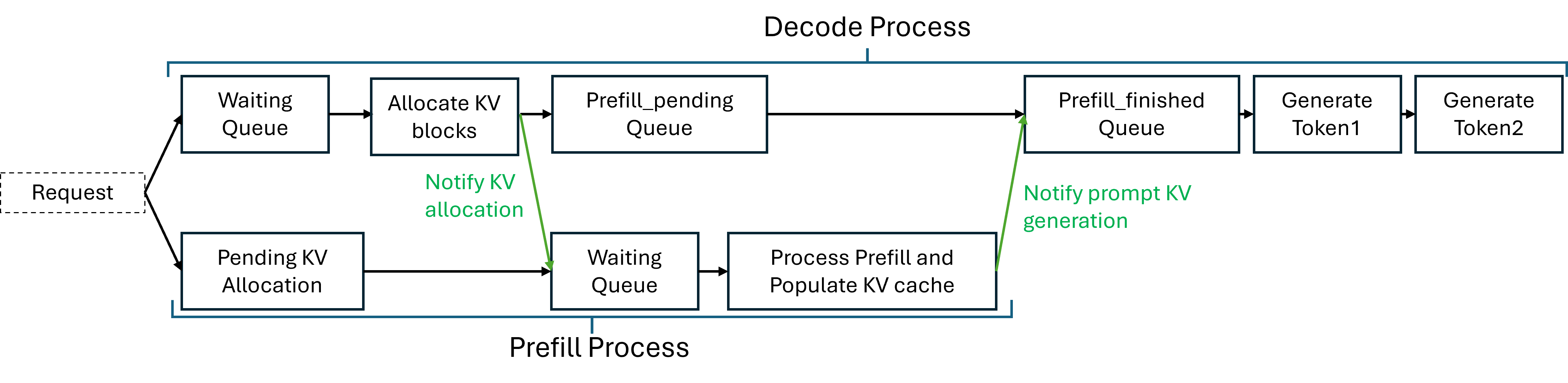}
    \caption{Flow of a single request}
    \label{fig:req_flow}
    \end{center}
\end{figure*}
\subsection{Request Flow}
Figure~\ref{fig:req_flow} shows the execution flow of a single request as it arrives in the RAPID-Serve engine. Each request is directed simultaneously to both prefill and decode processes. In the decode process, the request is added first to the "waiting queue," after which it is 
allocated the required KV cache blocks. The number of blocks to allocate for prompt can be determined by the input context length. The IDs of the allocated blocks are then passed to the prefill instance. At this point in the prefill process, the request moves from "pending\_kv queue" to "waiting\_prefill queue." This request is serviced by the prefill instance. After the prompt is processed and the KV cache populated, the decode process is notified. Note that there is no KV cache transfer, just a notification from prefill process to decode process. As the decode process receives this notification, it moves the request to "prefill\_finished" queue and subsequently adds it to the running decode batch to start generating the output tokens. This flow is fully asynchronous as there are no locks or serialization between the two processes. As there are multiple requests in the system at any one point in time, prefill process does not need to pause its execution of ongoing requests to wait for notifications from decode process and vice versa. 

\subsection{Server Components}
\begin{figure}[htbp]
    \begin{center}
    \includegraphics[width=\linewidth]{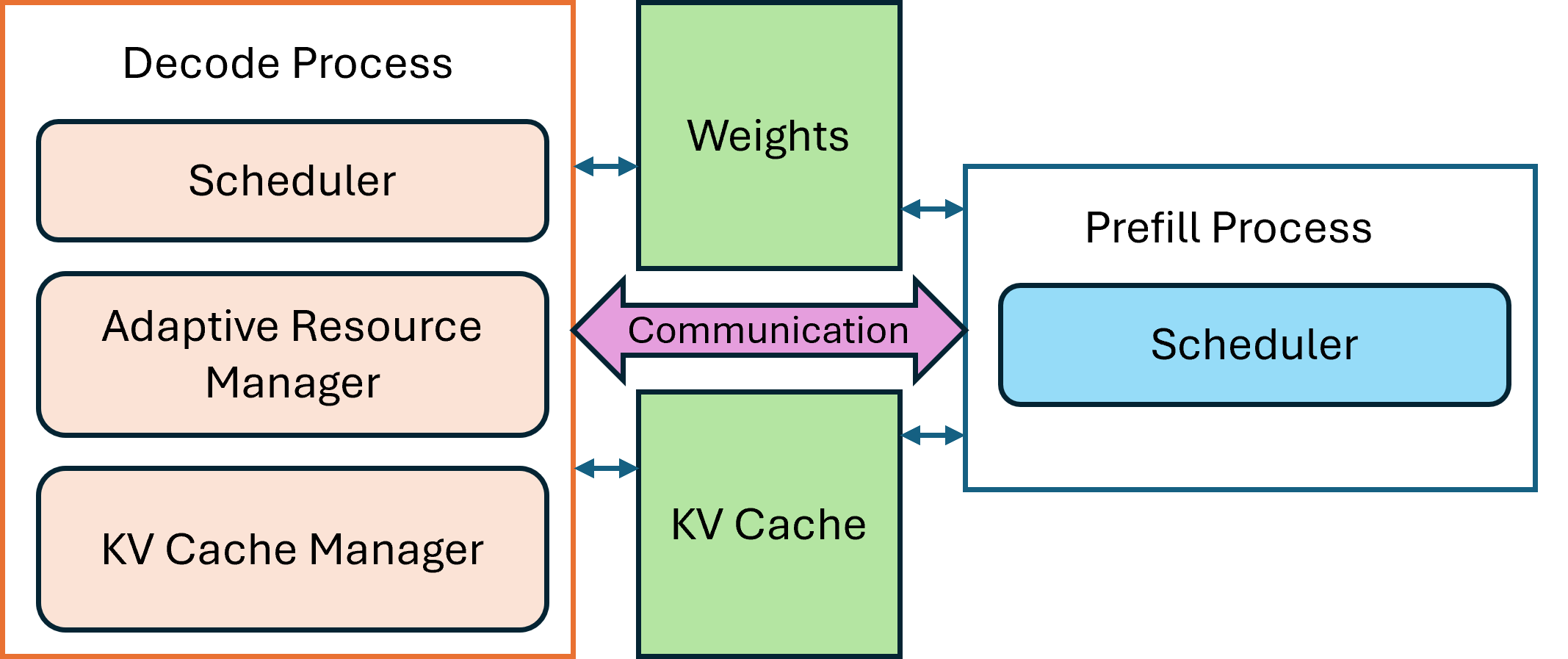}
    \caption{Main components of RAPID-Serve}
    \label{fig:server_comp}
    \end{center}
\end{figure}
The RAPID-Serve engine has two concurrent processes: prefill and decode. The model weights and KV cache are shared by both processes through Inter Process Communication (IPC) handles. The processes communicate through shared storage to send notifications of certain events and other information. Figure~\ref{fig:server_comp} presents the main components of RAPID-Serve design and the following sections present the details of these components.  
\subsubsection{KV Cache Manager}
The KV cache manager for RAPID-Serve is based on the vLLM KV cache manager that supports PagedAttention~\cite{pagedattention} to efficiently manage KV cache "block" allocation per request. It is also responsible for releasing the unused blocks as the request life-cycle ends or the request is preempted. In RAPID-Serve, only the decode process has this component. The reason is to avoid having both processes manage the KV cache simultaneously, which would require all KV cache operations to be done inside critical sections managed by locks, introducing additional serialization and dependencies between the two processes and violating design goal \#2. Hence, only the decode instance is responsible for managing the KV cache, because while the number of KV-cache blocks required for the prompt can be calculated based on the input context length, this calculation is not possible for the generated text (decode) as the number of output tokens are not known beforehand. As depicted in Figure~\ref{fig:req_flow}, the decode process allocates the KV cache blocks for prompt and notifies prefill, after which prefill process can start execution, avoiding the use of locks.

\subsubsection{Scheduler}
\begin{figure*}[htbp]
    \centering
    \begin{subfigure}[t]{0.33\textwidth}
    \includegraphics[height=4.5cm,width=!]{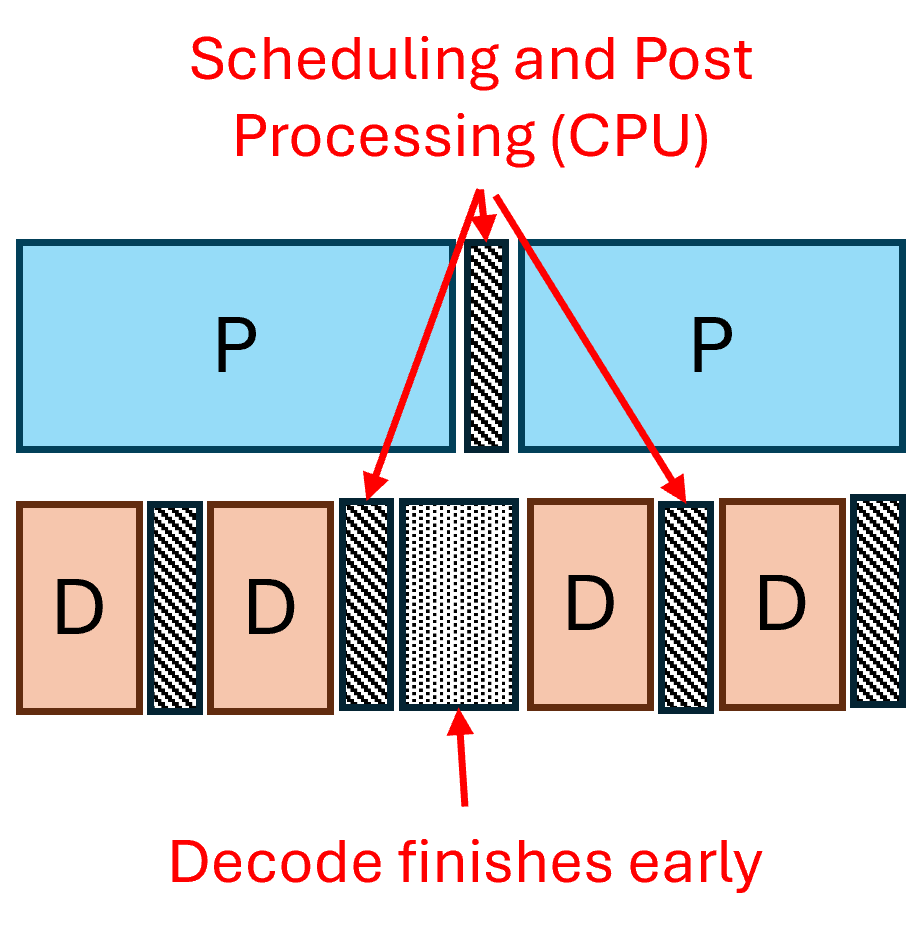}
    \caption{Naive scheduling and resource allocation}
    \label{naive}
    \end{subfigure}
    \hfill
    \begin{subfigure}[t]{0.33\textwidth}
    \includegraphics[height=4.5cm,width=!]{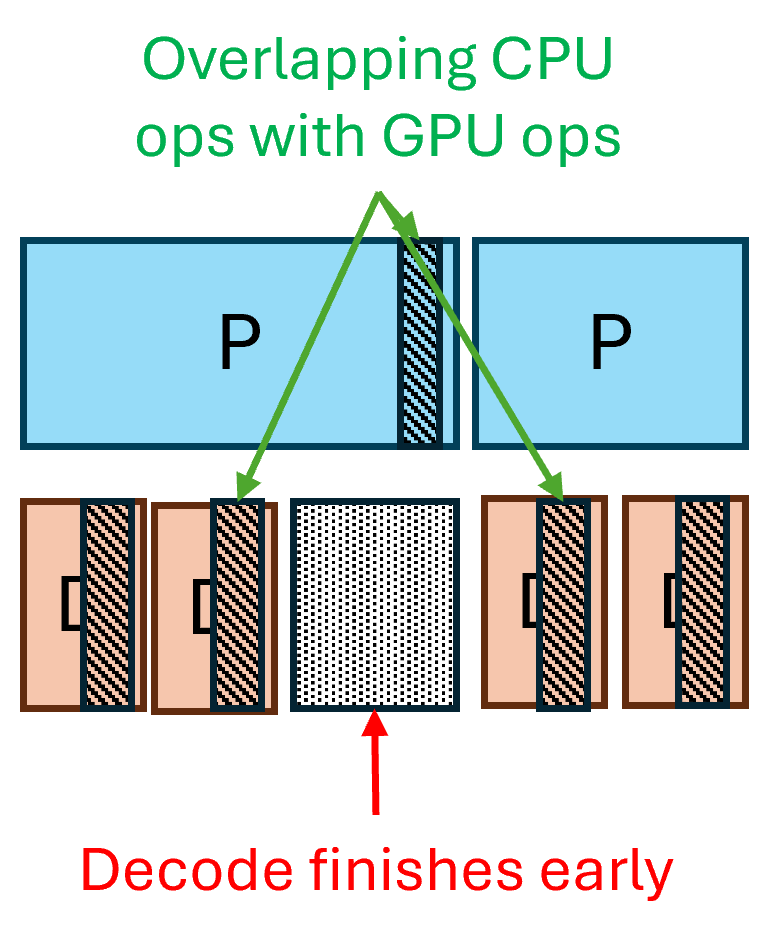}
    \caption{Async\_scheduling}
    \label{async}
    \end{subfigure}
    \hfill
    \begin{subfigure}[t]{0.33\textwidth}
    \includegraphics[height=4.5cm,width=!]{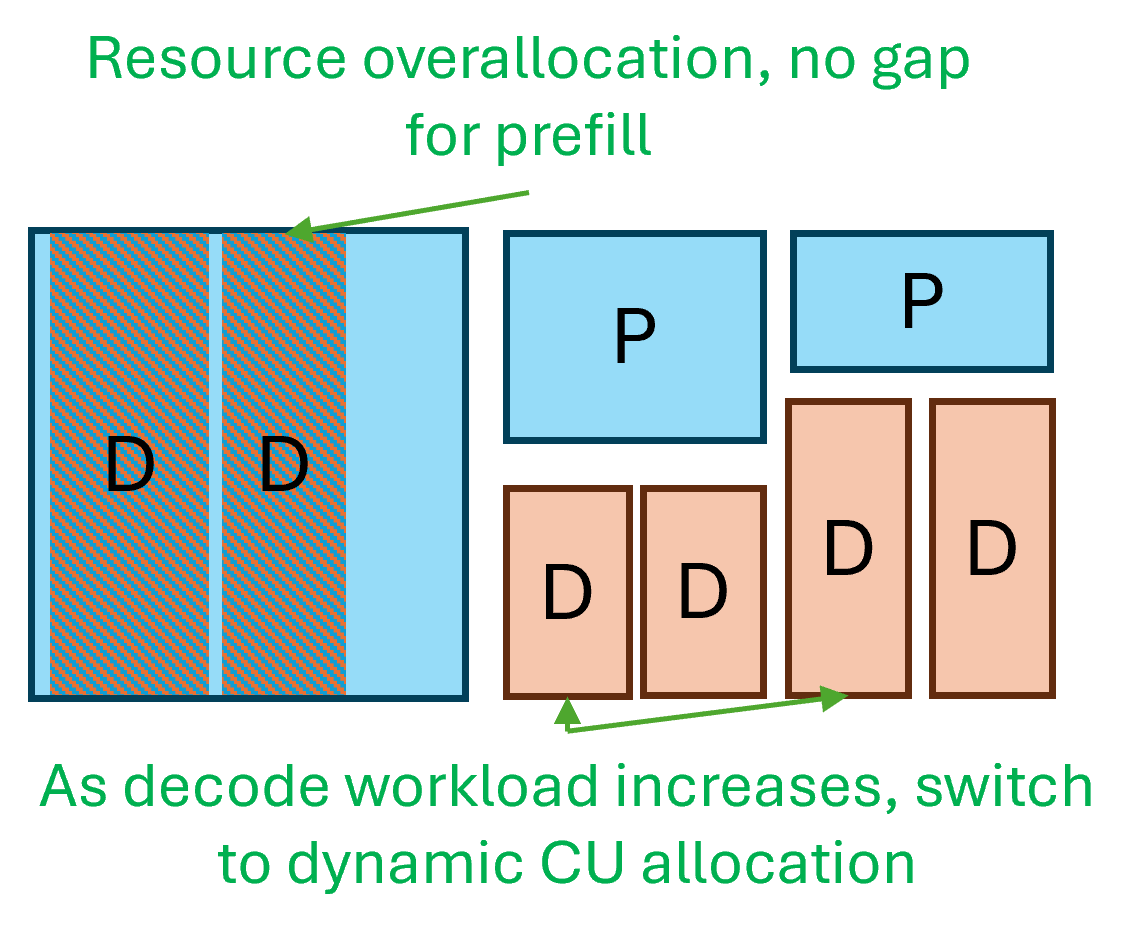}
    \caption{Adaptive Resource Allocation (CPU work not shown)
    }
    \label{final}
    \end{subfigure}
    \caption{Scheduling and Resource Allocation Strategies employed in RAPID-Serve}
    \label{scheduling}
\end{figure*}

The RAPID-Serve schedulers (both prefill and decode) follow a first-come-first-serve (FCFS) policy (i.e., requests are prioritized in the order of the arrival). 
To increase GPU utilization and reduce gaps due to scheduling (Figure~\ref{naive}), we enable async\_scheduling (Figure~\ref{async}).
This technique was introduced in Nanoflow~\cite{nanoflow2024} where the scheduler runs one step ahead of the GPU execution, considering the next token as a placeholder. If the request ends in the current step, an additional token would be generated in the next step since the async\_scheduler does not know that the request had ended at the time of scheduling (while the current step was still in flight). However, this is not a significant overhead as the decode requests are batched and the impact of one extra token per request is not significant. Async\_scheduling significantly decreases CPU time, specifically for the decode instance as the CPU time in decode is relatively higher because of lower GPU execution time. Async\_scheduling is also enabled for prefill and does not have the overhead of an additional generated token, as it is assumed that prefill request will finish in one step (no prefill chunking). Overall, async\_scheduling reduces the gaps in GPU execution and helps maintain high resource utilization.

\subsubsection{Adaptive Resource Manager}
The Adaptive Resource Manager is responsible for allocating compute resources based on current worklaod and SLOs. As described earlier, RAPID-Serve uses CU masking to allocate distinct resources to prefill and decode. However, when the number of decode requests is low, situations may arise when decode might finish early while prefill is still running and the CUs assigned to decode would remain unused during the current prefill iteration (Figure~\ref{async}). As RAPID-Serve makes use of HIP Graphs
for reducing the launch overhead of GPU kernels, it is not possible to change the command queue or its associated CU mask once the graph has been launched. Note that the opposite situation (i.e., prefill iteration ending earlier than decode) is not much of a concern since decode iteration time is much lower than prefill's, and all CUs can be assigned to decode on the next step.

To avoid this situation, the Adaptive Resource Manager overallocates the compute units and assigns 100\% of compute resources to both prefill and decode when decode workload is lower. The prefill and decode kernels end up competing for resources and have to rely on the GPU's hardware scheduler for their workgroups to be scheduled on any available compute unit. This scheme maximizes GPU utilization as 
the hardware scheduler can allocate any workgroup on any CU without any restriction. In case the decode ends while prefill iteration is still ongoing, the prefill workgroups will be assigned to all the CUs, mitigating the gap as shown in Figure~\ref{final}. 

\begin{figure}[htbp]
    \begin{center}
    \includegraphics[width=0.9\linewidth]{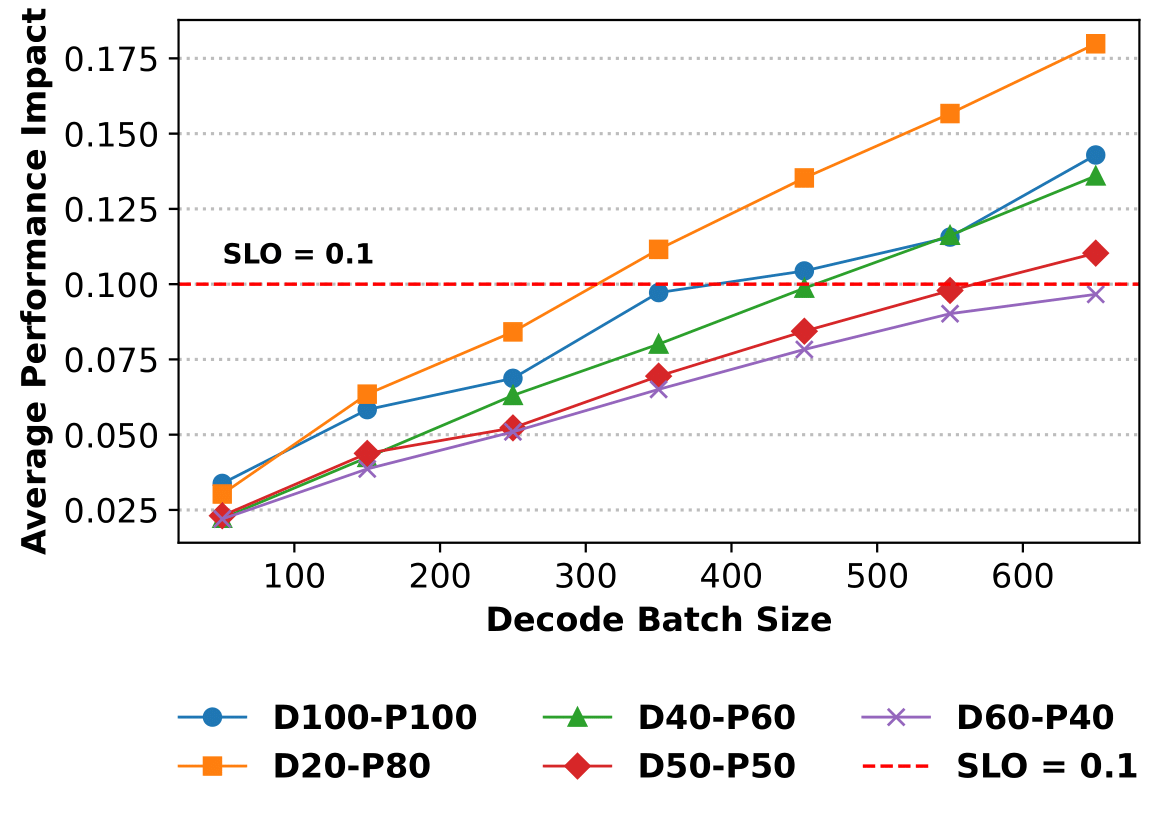}
    \caption{Decode latency at varying batchsizes with different resource allocation schemes. DXX=XX\% of CUs allocated to decode, PYY=YY\% of CUs allocated to prefill}
    \label{fig:decode_interference}
    \end{center}
\end{figure}
At higher request rates or heavier workload, decode requests fail to meet SLOs under the overallocation scheme.
This is because the competition between prefill and decode for the same resources introduces inter-stream interference as measured in Figure~\ref{fig:decode_interference}. The configuration P100-D100 (100\% to prefill, 100\% to decode) represents the overallocation scenario, which starts exceeding the SLO for larger decode batch sizes. 
It is also shown that at higher batch sizes, allocating distinct compute resources to prefill and decode eliminates inter-stream interference, allowing decode to meet the SLOs. 

Based on the observations in Figure~\ref{fig:decode_interference}, the Adaptive Resource Scheduler switches to distinct CU allocation schemes based on the decode workload and stringent ITL SLO. Based on offline profiles, the minimum number of compute units required for different decode batch sizes to meet the SLO are determined. At runtime, the manager allocates only enough CUs to decode as required to meet the SLO and to avoid impacting prefill performance more than needed as compute-bound prefill will get the remaining CUs and its performance will be impacted accordingly.
As the decode workload increases/decreases, the resource manager is able to adjust the number of resources to 
maintain high throughput and CU utilization while meeting SLOs.



\section{Evaluation}
\subsection{Experiment Setup}
All experiments were conducted on a single node with 8 AMD Instinct\textsuperscript{TM} MI300X GPUs with ROCm 6.4. The baseline serving system is vLLM v0.10.2. Two LLMs are evaluated: LlaMA-3 70B to represent dense transformer workloads, and Mixtral 8×7B to represent mixture-of-experts workloads. Three representative datasets are used for comprehensive evaluation: LMSYS~\cite{lmsys2023chatbotarena} trace has shorter prompts, which reflects interactive request patterns observed in real-world deployments, with an average prompt size of 2000 tokens. For longer-context workloads, the arXiV~\cite{arxivdataset} (average prompt size=8K tokens) and Loogle summarization~\cite{loogle2023} (average prompt size=20K) datasets are used which stress memory pressure. These datasets were truncated to 10K datapoints for LMSYS and arXiV datasets with stratified sampling by prompt length.
Experiments sweep load from under-utilized to saturated conditions with varying input Queries per second (QPS). RAPID-Serve performance is compared against hybrid batching at three different chunk sizes, as the performance of hybrid batching varies with chunk sizes. For disaggregated serving, both prefill and decode instances have tensor parallelism of 4 (4 GPUs each) and is also based on vLLM v1. Mixtral 8x7B model has not been evaluated in a disaggregated setup as vLLM currently does not support MoE models for disaggregation. 
\subsection{Throughput and Goodput}
\begin{figure*}[htbp]
  \centering
  \includegraphics[width=\linewidth]{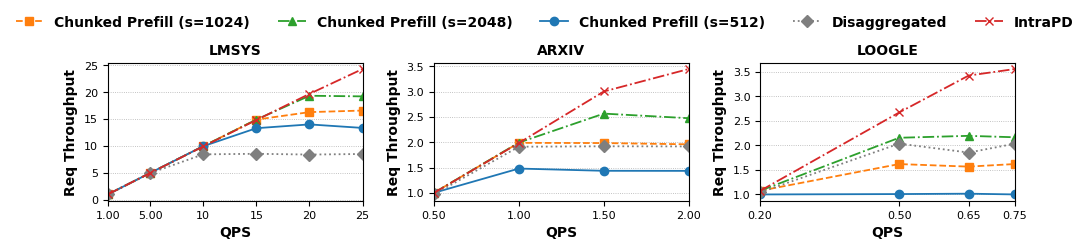}


  \includegraphics[width=0.9\linewidth]{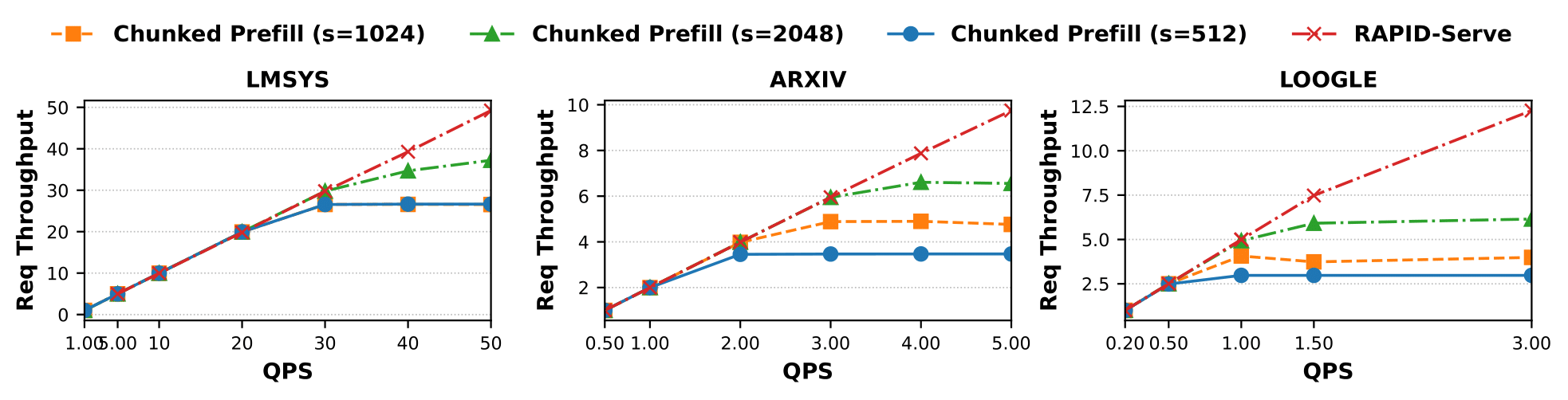}

  \caption{
  Throughput across workloads for LlaMA-3.1 70B (top) and Mixtral-8$\times$7B (bottom).
  Each plot shows offered load (QPS) versus achieved throughput under three prefill chunk sizes (hybrid batching),
  disaggregation and RAPID-Serve. Results are normalized to chunked (s=512) achieved throughput at the lowest QPS.
  }
  \label{fig:throughput}
\end{figure*}

\begin{figure*}[htbp]
  \centering
  \includegraphics[width=\linewidth]{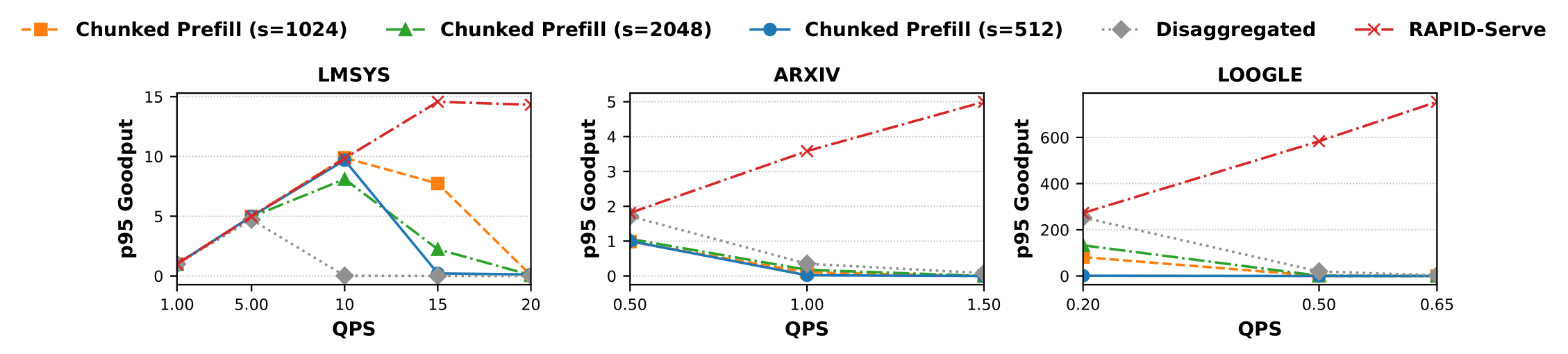}


  \includegraphics[width=0.9\linewidth]{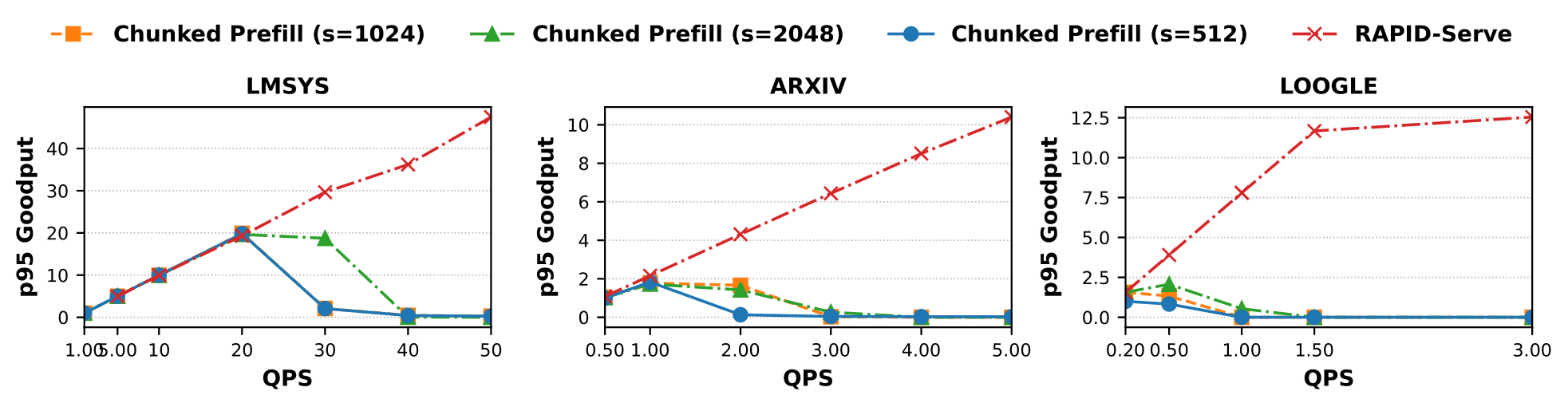}

  \caption{
  Goodput across workloads for LlaMA-3.1 70B (top) and Mixtral-8$\times$7B (bottom).
  Each plot shows offered load (QPS) versus achieved goodput under three prefill chunk sizes,
  disaggregation and RAPID-Serve. Results are normalized to chunked (s=512) achieved goodput at the lowest QPS.
  }
  \label{fig:total_goodput}
\end{figure*}



Figure~\ref{fig:throughput} shows RAPID-Serve's throughput at different request loads and compares it with hybrid batching and disaggregated serving.
All results are normalized to the throughput of hybrid batching with chunk size of 512 tokens at the lowest input QPS. 
Results show that RAPID-serve achieves equivalent or better unconstrained throughput compared to baseline, up to 4.1x and average 1.7x over chunked hybrid batching, and up to 2.9x and average 1.5x over disaggregated serving. This is because RAPID-Serve enables high concurrency; reducing gaps and ensuring high compute utilization.

Goodput is defined as the "maximum request rate that can be served adhering to the SLO attainment goal"~\cite{zhong2024distserve}. A request is considered successful if it satisfies both latency constraints: First, its ITL must not exceed the SLO (100 ms for LlaMA-3.1 70B, 50ms for Mixtral 8x7B). Second, its TTFT must remain within a length-dependent ceiling: requests with input context of 0–1000 tokens must complete within 1 second, those with 1000–2000 tokens within 2 seconds, and proportionally thereafter. 

Figure~\ref{fig:total_goodput} shows that RAPID-Serve consistently achieves equivalent or better p95 goodput until the saturation point, which is the request load at which GPUs are maximally utilized, and further increasing the workload only introduces scheduling delays (mainly impacting TTFT). It should be noted from Figure~\ref{fig:total_goodput} that at higher request loads even before the saturation points, the baselines fail to meet SLOs (\(goodput\approx0\)) while RAPID-Serve continues to achieve high goodput.
In cases where the baseline goodput is not negligible, RAPID-Serve achieves goodput improvements of up to 32x (average 4.9x) over chunked hybrid batching and up to 13.7x (average 4.7x) over disaggregated serving.


\begin{figure*}[htbp]
  \centering
  \includegraphics[width=1\linewidth]{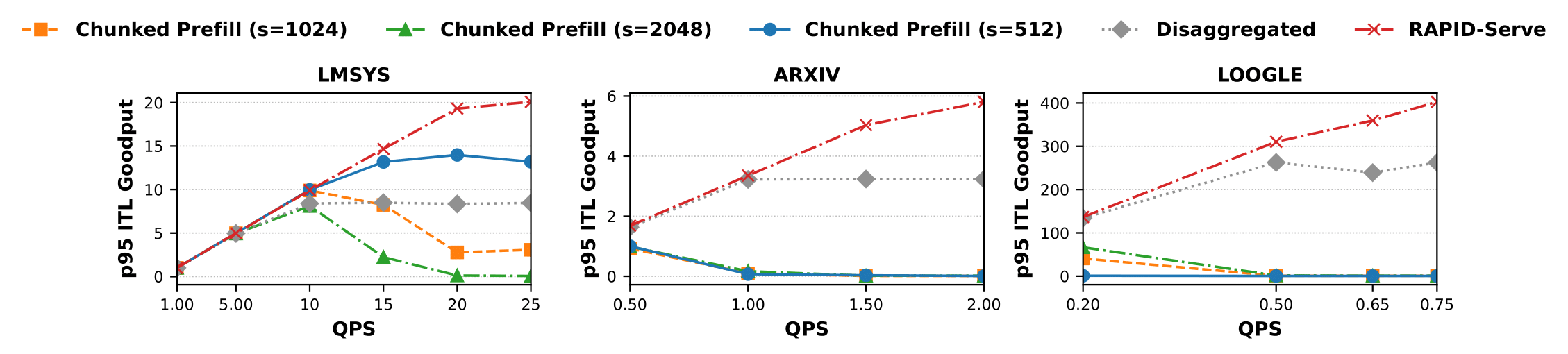}


  \includegraphics[width=0.9\linewidth]{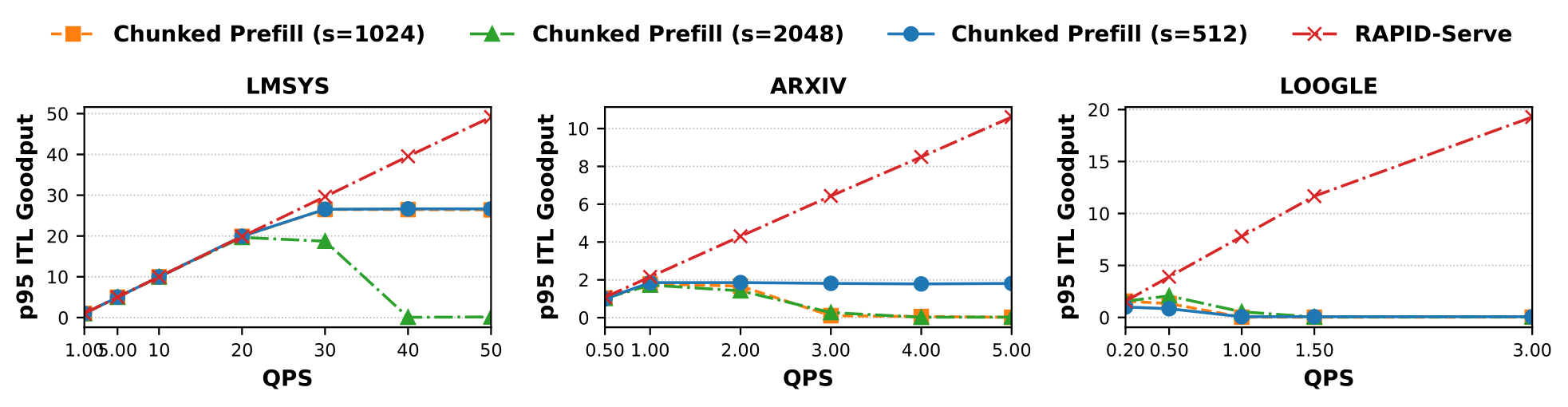}

  \caption{
  ITL p95 goodput across workloads for LlaMA-3.1 70B (top) and Mixtral-8$\times$7B (bottom).
  Each plot shows offered load versus achieved ITL p95 goodput under three prefill chunk sizes,
  the disaggregated baseline, and RAPID-Serve.
  Results are normalized to chunked (s=512) achieved goodput at the lowest QPS.
  }
  \label{fig:tpot_goodput}
\end{figure*}
As the TTFT gets longer due to queueing and scheduling delays after the saturation points, we also measure goodput by applying only the ITL SLO constraint (Figure~\ref{fig:tpot_goodput}). Only considering the cases where the baseline goodput is not negligible, RAPID-Serve achieves ITL goodput improvement of up to 5.9x (average 2x) over chunked hybrid batching, and up to 2.4x (average 1.4x) over disaggregated serving. In the rest of the cases, RAPID-Serve continues to achieve high goodput even when the baseline fails to meet SLOs. This is due to the RAPID-Serve's Adaptive Resource Management policies that allocate distinct resources to P/D to meet ITL criteria while maintaining high throughput. 

\subsection{Tail Latencies}

\begin{figure*}[htbp]
  \centering
  \includegraphics[width=0.95\linewidth]{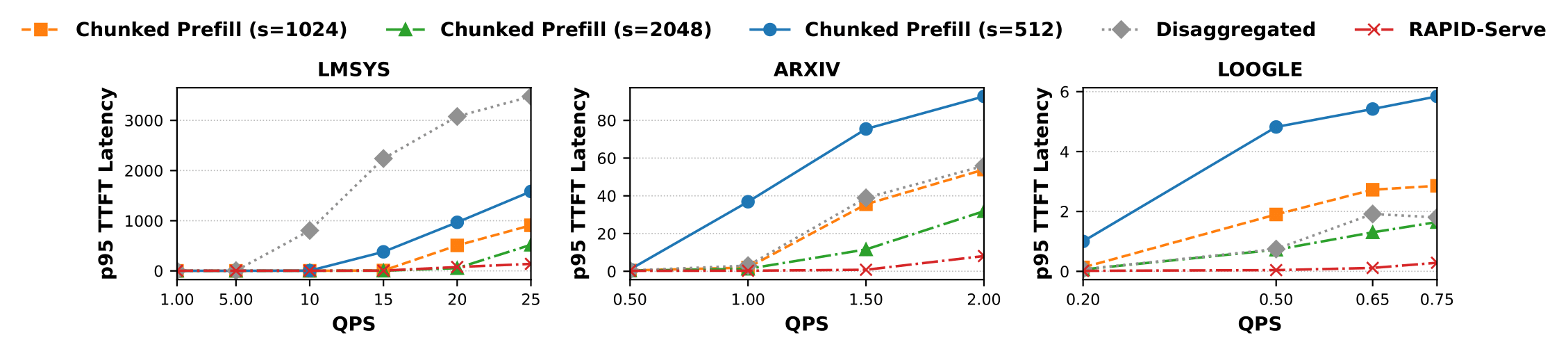}


  \includegraphics[width=0.95\linewidth]{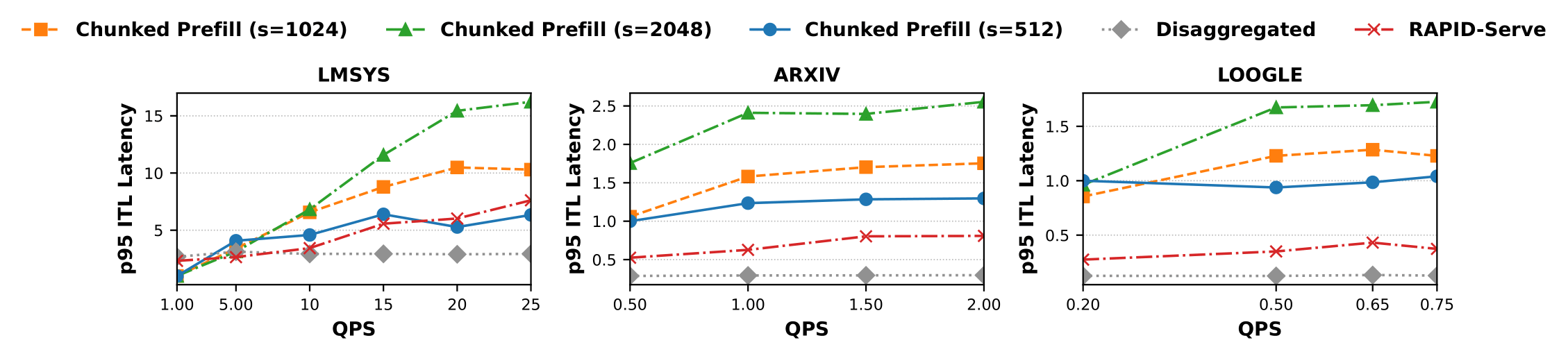}

  \caption{
  P95 TTFT and ITL for LlaMA-3.1 70B. Normalized to chunked (s=512) latencies at lowest QPS.
  }
  \label{fig:llama_tail}
\end{figure*}

RAPID-Serve strives to provide a balance between TTFT, ITL and throughput. An important metric to judge the SLO attainment of requests is tail latencies. As shown in Figure~\ref{fig:llama_tail}, the p95 TTFT latency for LlaMA-3.1 70B model is up to 220x (average 53x) and 486x (average 55x) lower than chunked and disaggregated baselines respectively. This improvement in TTFT is due to the following factors: 1) Prefill is not chunked in RAPID-Serve hence there is no repeated overhead of scheduling, model loading, kernel launches etc., and 2) there is no KV cache transfer during the first token generation. 

The p95 ITL latency is up to 6x (average 1.9x) lower than chunked hybrid batching. Disaggregation has on average 2x lower p95 ITL than RAPID-Serve, however this lower ITL can be attributed to disaggregation having lower throughput at the same time. 
\subsection{Resource Utilization}
RAPID-Serve improves GPU resource utilization through concurrent prefill and decode execution, asynchronous scheduling, and a lock-free design that mitigates GPU idle gaps. Compute utilization is measured using hardware performance counters: the ratio of the sum of busy cycles per compute unit and the number of busy GPU cycles is used as 
an approximation of active compute units. RAPID-Serve improves GPU compute utilization by up to 77\% compared to hybrid batching and 111\% compared to disaggregated serving. Moreover, RAPID-Serve avoids memory utilization imbalance of disaggregated serving by sharing KV cache, improving memory utilization by up to 37\%.

\begin{figure*}[htbp]
  \centering
  \includegraphics[width=0.9\linewidth]{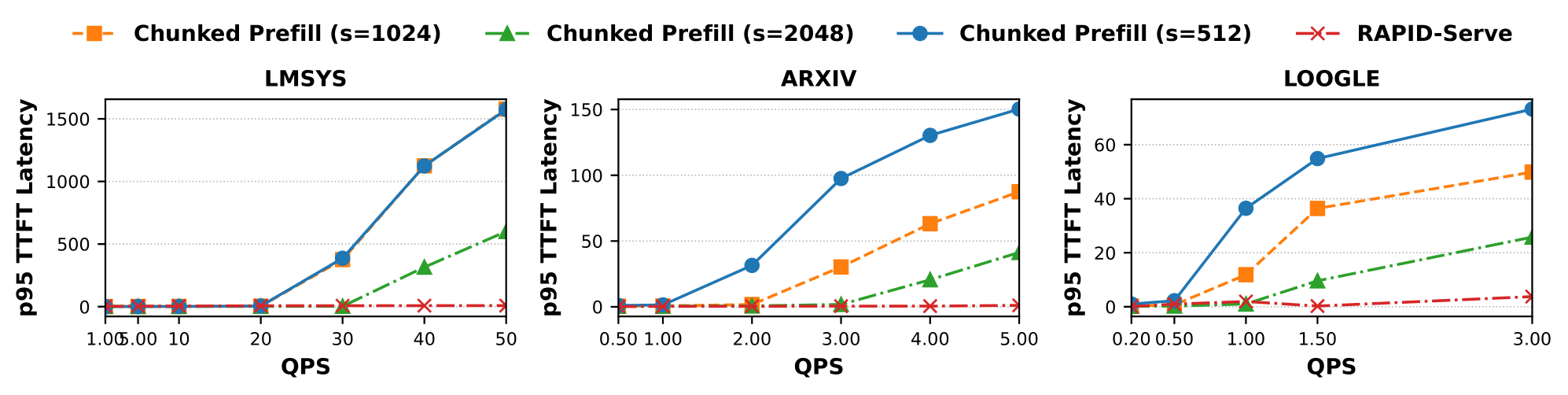}


  \includegraphics[width=0.9\linewidth]{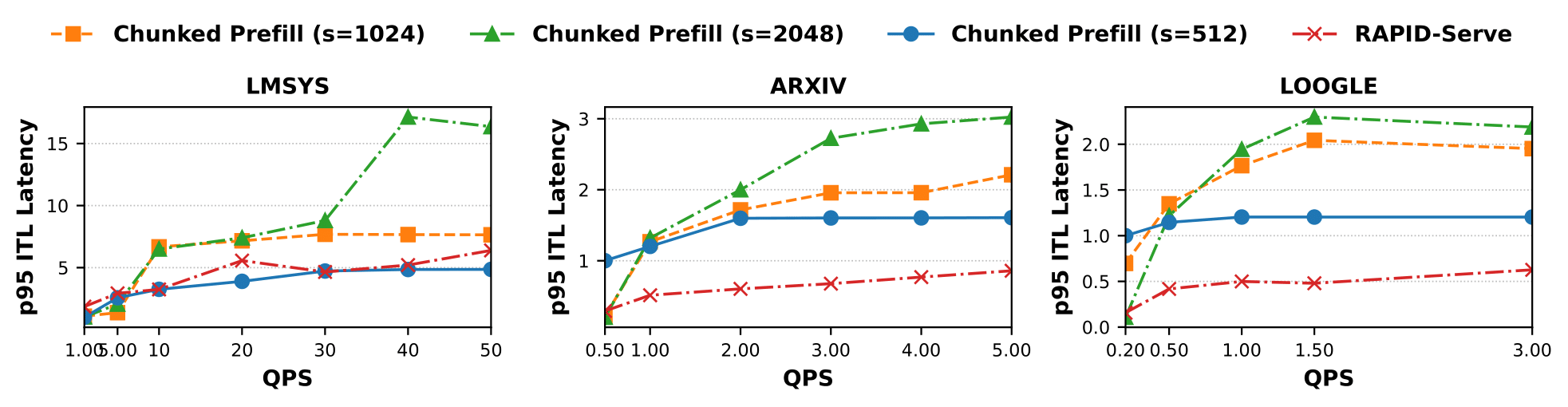}

  \caption{
  P95 TTFT and ITL for Mixtral8x7B. Normalized to chunked (s=512) latencies at lowest QPS.
  }
  \label{fig:mixtral_tail}
\end{figure*}
\section{Related Work}
There have been several approaches to addressing the issues faced by current serving systems. PODAttention~\cite{podattention2024} presents a unified attention kernel that combines both prefill and decode attention. However, their approach is specific to hybrid batching and does not resolve the issues inherent to the serving approach as discussed in this work. Nanoflow~\cite{nanoflow2024} takes a different approach to overlapping kernels and improving GPU utilization. Each batch is divided into multiple "nano-batches" and executed concurrently, improving resource utilization. However, P/D are still coupled together and Nanoflow does not address the resulting per-token latency degradation.

DRIFT~\cite{drift} and Semi-PD~\cite{semipd} overlap prefill and decode on the same devices. DRIFT utilizes NVIDIA's GreenContext mechanism for resource allocation, while SemiPD achieves this through NVIDIA MPS. These approaches allocate distinct resources to prefill and decode, resulting in bubbles during execution. RAPID-Serve addresses this through overallocation of resources depending on the workload, ensuring that when either decode or prefill is not running, the GPU is not underutilized. DRIFT only evaluates ITL SLO constraints while RAPID-Serve's Adaptive Resource scheduling provides a balance between prefill and decode meeting both TTFT and ITL constraints. Moreover, RAPID-Serve's lock and sync-free approach mitigates serialization and synchronization between prefill and decode and improves performance and throughput.

\section{Conclusions}
RAPID-Serve provides an effective solution to address performance limitations of current LLM inference serving solutions, providing high throughput while meeting the SLO constraints. Instead of coupling prefill and decode requests together as in hybrid batching, which can result in requests exceeding the latency SLOs, RAPID-Serve overlaps execution of prefill and decode without combining them in the same phase. This enables RAPID-Serve to  maintain high throughput while meeting the SLOs. Moreover, RAPID-Serve also avoids the KV cache transfer overhead and memory capacity underutilization of disaggregated serving. Evaluations show that RAPID-Serve provides significant performance uplifts: up to 4.1x unconstrained throughput and 32x goodput improvements. These results show the efficacy of concurrent execution of prefill and decode phases to efficiently utilize the available GPU resources.



\bibliographystyle{plain}
\bibliography{main}

\begin{thebibliography}{10}

\bibitem{agrawal2024taming}
Amey Agrawal, Nitin Kedia, Ashish Panwar, Jayashree Mohan, Nipun Kwatra, Bhargav~S. Gulavani, Alexey Tumanov, and Ramachandran Ramjee.
\newblock Taming throughput-latency tradeoff in llm inference with sarathi-serve.
\newblock In {\em Proceedings of the 18th USENIX Symposium on Operating Systems Design and Implementation (OSDI ’24)}, 2024.

\bibitem{agrawal2024medha}
Amey Agrawal, Haoran Qiu, Junda Chen, Íñigo Goiri, Chaojie Zhang, Rayyan Shahid, Ramachandran Ramjee, Alexey Tumanov, and Esha Choukse.
\newblock Medha: Efficiently serving multi-million context length llm inference requests without approximations.
\newblock {\em arXiv preprint arXiv:2409.17264}, 2024.
\newblock Available at \url{https://arxiv.org/abs/2409.17264}.

\bibitem{chen2025_disaggregated_inference_18mo_later}
Junda Chen, Yonghao Zhuang, and Hao Zhang.
\newblock Disaggregated inference: 18 months later.
\newblock \url{https://hao-ai-lab.github.io/blogs/distserve-retro/}, November 2025.
\newblock Accessed: 2025-11-29.

\bibitem{Cheng2025LMCache}
Yihua Cheng, Yuhan Liu, Jiayi Yao, Yuwei An, Xiaokun Chen, Shaoting Feng, Yuyang Huang, Samuel Shen, Kuntai Du, and Junchen Jiang.
\newblock Lmcache: An efficient kv cache layer for enterprise-scale llm inference.
\newblock {\em CoRR}, abs/2510.09665, 2025.

\bibitem{arxivdataset}
Arman Cohan, Franck Dernoncourt, Doo~Soon Kim, Trung Bui, Seokhwan Kim, Walter Chang, and Nazli Goharian.
\newblock A discourse-aware attention model for abstractive summarization of long documents.
\newblock In {\em Proceedings of the 2018 Conference of the North {A}merican Chapter of the Association for Computational Linguistics: Human Language Technologies, Volume 2 (Short Papers)}, pages 615--621, New Orleans, Louisiana, June 2018. Association for Computational Linguistics.

\bibitem{drift}
Weihao Cui, Yukang Chen, Han Zhao, Ziyi Xu, Quan Chen, Xusheng Chen, Yangjie Zhou, Shixuan Sun, and Minyi Guo.
\newblock Optimizing slo-oriented llm serving with pd-multiplexing, 2025.

\bibitem{deepspeedfastgen2024}
Christian Holmes, Masahiro Tanaka, Matthew Wyatt, A.~A. Awan, Jeff Rasley, Samyam Rajbhandari, others, and Yuxiong He.
\newblock Deepspeed-fastgen: High-throughput text generation for llms via mii and deepspeed-inference.
\newblock {\em arXiv preprint arXiv:2401.08671}, 2024.

\bibitem{semipd}
Ke~Hong, Lufang Chen, Zhong Wang, Xiuhong Li, Qiuli Mao, Jianping Ma, Chao Xiong, Guanyu Wu, Buhe Han, Guohao Dai, Yun Liang, and Yu~Wang.
\newblock semi-pd: Towards efficient llm serving via phase-wise disaggregated computation and unified storage, 2025.

\bibitem{podattention2024}
Abhijith~Kalyan Kamath, Rakesh Prabhu, Jatin Mohan, Sajan Peter, Radhika Ramjee, and Animesh Panwar.
\newblock Pod-attention: Unlocking full prefill-decode overlap for faster llm inference.
\newblock {\em arXiv preprint arXiv:2410.18038}, 2024.

\bibitem{pagedattention}
Woosuk Kwon, Zhuohan Li, Siyuan Zhuang, Ying Sheng, Lianmin Zheng, Cody~Hao Yu, Joseph~E. Gonzalez, Hao Zhang, and Ion Stoica.
\newblock Efficient memory management for large language model serving with pagedattention, 2023.

\bibitem{loogle2023}
Xinyu Li, Lianwei Wu, Haotian Zhang, et~al.
\newblock Loogle: A long-context dataset for large language model evaluation, 2023.

\bibitem{mitra2025beyondbuzz}
Tiyasa Mitra, Ritika Borkar, Nidhi Bhatia, Ramon Matas, Shivam Raj, Dheevatsa Mudigere, Ritchie Zhao, Maximilian Golub, Arpan Dutta, Sailaja Madduri, Dharmesh Jani, Brian Pharris, and Bita Darvish~Rouhani.
\newblock Beyond the buzz: A pragmatic take on inference disaggregation.
\newblock {\em arXiv preprint arXiv:2506.05508}, 2025.

\bibitem{fastertransformer}
{NVIDIA}.
\newblock Fastertransformer.
\newblock \url{https://github.com/NVIDIA/FasterTransformer}, 2024.

\bibitem{splitwise2024}
Parijat Patel, Ekta Choukse, Chi Zhang, Aman Shah, Inigo Goiri, Saeed Maleki, and Ricardo Bianchini.
\newblock Splitwise: Efficient generative llm inference using phase splitting.
\newblock In {\em 2024 ACM/IEEE 51st Annual International Symposium on Computer Architecture (ISCA)}, pages 118--132. IEEE, 2024.

\bibitem{Qin2025Mooncake}
Ruoyu Qin, Zheming Li, Weiran He, Jialei Cui, Feng Ren, Mingxing Zhang, Yongwei Wu, Weimin Zheng, and Xinran Xu.
\newblock Mooncake: Trading more storage for less computation — a kvcache-centric architecture for serving {LLM} chatbot.
\newblock In {\em 23rd USENIX Conference on File and Storage Technologies (FAST 25)}, pages 155--170, Santa Clara, CA, February 2025. USENIX Association.

\bibitem{yu2022orca}
Gyeong-In Yu, Joo~Seong Jeong, Geon-Woo Kim, Soojeong Kim, and Byung-Gon Chun.
\newblock Orca: A distributed serving system for transformer-based generative models.
\newblock In {\em 16th USENIX Symposium on Operating Systems Design and Implementation (OSDI ’22)}, pages 521--538, Carlsbad, CA, 2022. USENIX Association.

\bibitem{lmsys2023chatbotarena}
Lianmin Zheng, Wei-Lin Chiang, Ying Sheng, Sodaice Zhuang, Ztao Wu, Zhe Lin, et~al.
\newblock Chatbot arena: An open platform for evaluating llms by human preference, 2023.

\bibitem{zhong2024distserve}
Yinmin Zhong, Shengyu Liu, Junda Chen, Jianbo Hu, Yibo Zhu, Xuanzhe Liu, Xin Jin, and Hao Zhang.
\newblock Distserve: Disaggregating prefill and decoding for goodput-optimized large language model serving.
\newblock In {\em Proceedings of the 18th USENIX Symposium on Operating Systems Design and Implementation (OSDI ’24)}, 2024.

\bibitem{nanoflow2024}
Kaiyuan Zhu, Yizhou Zhao, Li~Zhao, Guangming Zuo, Yongbin Gu, Dongyan Xie, others, and Baris Kasikci.
\newblock Nanoflow: Towards optimal large language model serving throughput.
\newblock {\em arXiv preprint arXiv:2408.12757}, 2024.

\end{thebibliography}

\end{document}